\documentclass[aps,prl,twocolumn,superscriptaddress,showpacs]{revtex4}

\usepackage{graphicx}

\begin{document}


\title{Optical bistability involving planar metamaterial with broken structural symmetry}


\author{Vladimir R. Tuz}
\affiliation{Institute of Radio Astronomy of National Academy of
Sciences of Ukraine, 4, Krasnoznamennaya st., Kharkiv 61002,
Ukraine}
\affiliation{School of Radio Physics, Karazin Kharkiv
National University, 4, Svobody Square, Kharkiv 61077, Ukraine}

\author{Sergey L. Prosvirnin}
\affiliation{Institute of Radio Astronomy of National Academy of
Sciences of Ukraine, 4, Krasnoznamennaya st., Kharkiv 61002,
Ukraine}
\affiliation{School of Radio Physics, Karazin Kharkiv
National University, 4, Svobody Square, Kharkiv 61077, Ukraine}

\author{Lyudmila A. Kochetova}
\affiliation{Institute of Radio Astronomy of National Academy of Sciences of Ukraine, 4, Krasnoznamennaya st., Kharkiv 61002, Ukraine}



\begin{abstract}
We report on a bistable light transmission through a planar
metamaterial composed of a metal pattern of weakly asymmetric
elements placed on a nonlinear substrate. Such structure bears the
Fano-like sharp resonance response of a trapped-mode excitation. The
feedback required for bistability is provided by the coupling
between the strong antiphased trapped-mode-resonance  currents
excited on the metal elements and the intensity of inner field in
the nonlinear substrate.
\end{abstract}




\maketitle


The optical transmission bistability is a phenomenon whereby a
system changes its transmission from one value to another in
response to properties of light passing through \cite{gibs-1985}.
The switch between the two stable states of transmission is usually
induced by the intensity of the incident light, and, luckily, the
switch may exhibit a hysteresis. The light intensity driven systems
must have an intrinsic nonlinear susceptibility and feedback to
provide optical bistability. This effect can be used to realize
all-optical switches, limiters and logic gates. The issue of the day
is to reduce the size of such all-optical devices, decrease their
switching times and the required intensity of light.

Typically, all-optical devices include a Fabry-Perot resonator to
provide feedback \cite{felber-1976-mod}; although the systems
without such resonator can be used too. An example of such
structures is the nonlinear photonic crystal microresonators
\cite{soljacic-2002-obs,bravo-2007-eno} and the devices based on the
surface plasmon-polaritons in metal nanostructures
\cite{wurtz-2006-obn,min-2008-aos}. They offer a unique mechanism
for confining light, giving rise to a combination of high quality
factors and small modal volumes that is helpful to stronger
nonlinear effects and reduce the size of the all-optical devices.

There is another promising way to develop the small-size all-optical
switching devices without electromagnetic cavities and surface
plasmons field enhancement. Here the case in point is the planar
metamaterials (also known as metafilms) with active constituents.
Typically, these systems are surface structures which consist of
some metal or dielectric resonance elements arranged as a periodic
array and placed on a layer thin in comparison to the wavelength.
One way to obtain the nonlinear response of a planar metamaterial is
to introduce some nonlinear individual resonance elements in the
structure (see for instance \cite{klein-2006-shg}). Thus in
\cite{powell-2009-nem}, the elements are made nonlinear and tunable
via the insertion of diodes with a voltage-controlled capacitance.
However, in the optical range the manufacturing of such structures
is associated with considerable technological difficulties. Another
simpler way is to arrange the proper resonance elements on a
nonlinear substrate.

The main feature of the planar metamaterials, essential for the
optical switching applications, is a resonance character of their
transmission and reflection spectra. The usual resonance field
enhancement inside a planar metamaterial may be extremely enlarged
by involving structures which bear so-called trapped or dark modes
\cite{stockman-2001-lvd,yariv-2009-git}. The excitation of
high-quality-factor trapped mode resonances in planar
double-periodic structures with a broken symmetry was shown both
theoretically \cite{prosvirnin-01-mla,prosvirnin-2003-rcm} and
experimentally \cite{fedotov-2007-stm} in microwaves. In particular,
these typical peak-and-trough Fano spectral profile resonances are
excited in the periodic structure consisted of asymmetrically split
metal rings. A small asymmetry of the metal elements of such
structure results in the excitation of the strong mode of
anti-phased currents, which provides low radiation losses and
therefore high \textit{Q}-factor resonances. Recently, the trapped
mode resonances were investigated in similar planar structures in
the near-IR range \cite{khardikov-2010-tol}. It was shown that the
special choice of geometry parameters of the structure enables to
increase the \textit{Q}-factor of the trapped mode resonance by
several times in comparison with the ordinary plasmon-polariton
resonance. Such high-\textit{Q} resonance regime is promised to
observe a bistability in the near-IR range, if the structure is to
include a nonlinear material.

In this report, we propose and study an all-optical switching device
based on the bistability in a planar metamaterial made of complex
shaped resonance particles placed on a substrate of nonlinear
material in the regime of a trapped mode excitation.

\begin{figure}
\includegraphics[width=80mm]{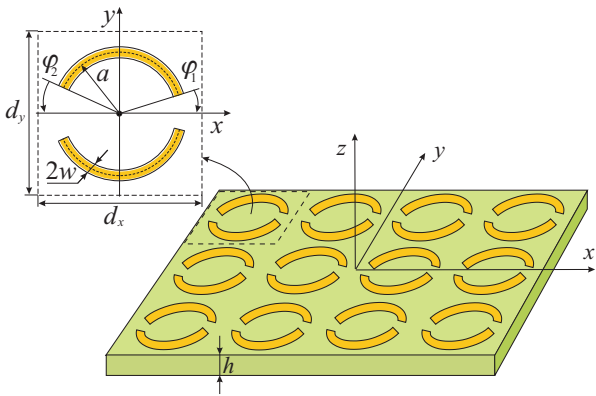}
\caption{(Color online) Fragment of the planar metamaterial and its
elementary cell. The size of the square translation cell is $d = d_x
= d_y = 900$~nm. The asymmetrically split metal rings are placed on
the top face of $h = 0.1d$ thick nonlinear dielectric layer. The
radius and the width of metal rings are $a = 0.4d$ and $2w = 0.06d$
respectively. The angle sizes of ring splits are $\varphi_1 =
15^\circ$ and $\varphi_2 = 25^\circ$.} \label{fig:structure}
\end{figure}

The studied structure consists of the identical subwavelength metal
inclusions in the form of asymmetrically split rings (ASRs) arranged
in a periodic array and placed on a thin nonlinear dielectric
substrate (see Fig.~\ref{fig:structure}). Each ASR contains two
identical strip elements opposite to one another. The right-hand
split between the strips $\varphi_1$ is a little different from the
left-hand one $\varphi_2$, so that the square unit cell is
asymmetric with regard to the $y$-axis. A normally incident plane
wave polarized transversely to the array symmetry axis ($x$-axis)
can excite the trapped mode resonances. Therefore we consider the
structure excitation by $y$-polarized plane wave. Suppose, that the
incident field has an amplitude $A$ and a frequency $\omega$.

As usual, actual optical array structures are mounted into a
dielectric host or placed on a thick dielectric substrate. Two
things happen with the trapped mode resonance frequency $\omega_0$,
when a dielectric substrate is added next to the structure. First,
the resonant frequency $\omega_0$ changes. If the structure is
completely inside the dielectric material with relative dielectric
constant $\varepsilon_r$, the resonant frequency will reduce in
frequency with the factor $\sqrt{\varepsilon_r}$ \cite{munk-2000}.
If the structure is placed between dielectric slabs of finite
thickness, the resonant frequency will change to somewhere between
$\omega_0$ and $\omega_0/\sqrt{\varepsilon_r}$. And, at last, if we
have a dielectric only to one side of the structure, the resonant
frequency shifts downward to be near
$\omega_0/\sqrt{(\varepsilon_r+1)/2}$. Second, the finite thickness
of a practical substrate (of about 0.3--0.5~mm) results in the
appearance of some additional interference resonances. However, as
shown in \cite{khardikov-2010-tol}, the trapped mode resonance can
be easily separated against the interference resonances background.
Thus the presence of the thick dielectric substrate has no effect on
the trapped mode resonance qualitatively; and we consider further a
"free standing" structure configuration of a metal array placed on a
thin nonlinear substrate, in order to simplify explanation.

The algorithm based on the method of moments was proposed earlier
\cite{prosvirnin-01-mla} to study the resonance nature of the
structure response, under the assumption of such a small amplitude
$A$ that the dependence of the substrate permittivity $\varepsilon$
on the field intensity is infinitesimal.

The algorithm requires that, at the first step, the surface current
induced in the metal pattern by the field of the incident wave is to
be calculated. The metal pattern is treated as a perfect conductor,
while the substrate is assumed to be a lossy dielectric. In
particular, assuming $A=1$~v/m the magnitude of the current $I$
averaged along the single element can be determined as a function
\begin{equation}\label{eq:one}
    I=Q(\omega, \varepsilon).
\end{equation}
At the second step, the found surface current distribution is used
to calculate the transmission and reflection coefficients as
$t=t(\omega, \varepsilon)$, $r=r(\omega, \varepsilon)$.

To introduce the nonlinearity (the third-order Kerr-effect), let us
assume that the permittivity $\varepsilon$ of the substrate depends
on the intensity of the electromagnetic field inside it. In our
approximate approach to the nonlinear problem solution, first assume
that the inner intensity is directly proportional to the square of
the current magnitude averaged over a metal pattern extent.
Secondly, in view of the smallness of the translation cell of the
array, we suppose that the nonlinear substrate remains to be a
homogeneous dielectric slab under intensive light. Thus, the
intensity-dependent permittivity of the substrate is given further
as
$$\varepsilon=\varepsilon_1+\varepsilon_2|{I}|^2.$$
Note, since the substrate permittivity is proportional to the
current value in the metal pattern, the nonlinearity effect reaches
its maximum under the ASR resonance condition.

If the amplitude $A$ of the incident field differs from the unity,
the appropriate average current magnitude for a given $\varepsilon$
can be found using (\ref{eq:one}) as
\begin{equation}\label{eq:four}
I=A \cdot Q(\omega, \varepsilon).
\end{equation}

Since the substrate permittivity $\varepsilon$ depends on the
average current value $I$, the relation (\ref{eq:four}) can be
rewritten as follows
\begin{equation}\label{eq:five}
I=A \cdot Q(\omega, \varepsilon_1+\varepsilon_2|{I}|^2).
\end{equation}

The expression (\ref{eq:five}) is a nonlinear equation related to
the average current value in the metal pattern. The incident field
magnitude is a parameter of the equation (\ref{eq:five}). At a fixed
frequency $\omega$, the solution of this equation is the average
current value dependent on the magnitude of the incident field
$I=I(A),$
where the function $I(A)$ is presumably multivalued.

On the basis of the current $I(A)$ found by a numerical solution of
the equation (\ref{eq:five}), it is possible now to determine the
permittivity of the nonlinear substrate
$\varepsilon=\varepsilon_1+\varepsilon_2|{I(A)}|^2$
and to calculate the reflection and transmission coefficients
$$t=t(\omega, \varepsilon_1+\varepsilon_2|{I(A)}|^2),\quad
r=r(\omega, \varepsilon_1+\varepsilon_2|{I(A)}|^2),$$
as the functions of the magnitude of the incident field.

\begin{figure}
\includegraphics[width=80mm]{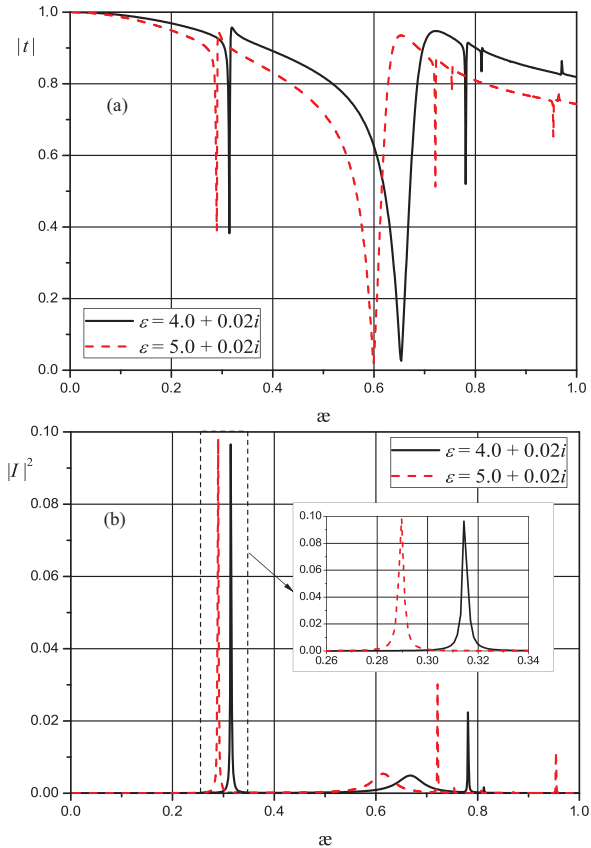}
\caption{(Color online) The frequency dependences of the magnitude
of the transmission coefficient (a) and the square of the current
magnitude (a.u.) averaged along the metal pattern (b) in the case of
the linear permittivity ($\varepsilon_2=0$) of the substrate.}
\label{fig:linearT&I}
\end{figure}

At first we consider the transmission through the array of ASRs
placed on a linear substrate (see Fig.~\ref{fig:linearT&I})
\cite{prosvirnin-2003-rcm,fedotov-2007-stm,khardikov-2010-tol}. If a
normal incident wave is polarized in $y$-direction, at the
dimensionless frequency nearly $\ae=d/\lambda\sim 0.3$, a sharp
reflection resonance occurs. This resonance corresponds to the
excitation of a trapped mode because equal and opposite directed
currents in the two arcs of each complex particle of array radiate a
little in free space. The resonance has a high quality factor, and
the current magnitude reaches the maximum at this frequency. As the
permittivity of the substrate increases, the resonance frequency
shifts to low values. Note, if the incident field is $x$-polarized
or the splits between the strips are the same, only a symmetric
current mode is excited. The corresponding resonance has a low
quality factor, and this configuration is not suitable to observe a
bistability because the current magnitude is small at this
frequency.

Suppose that the trapped mode resonance frequency is slightly higher
than the incident field frequency. As the intensity of the incident
field rises, the magnitude of currents on the metal elements
increases. This leads to increasing the field strength inside the
substrate and its permittivity as well. As a result, the frequency
of the resonant mode decreases and shifts toward the frequency of
incident wave, which, in turn, enhances further the coupling between
the current modes and the inner field intensity in the nonlinear
substrate. This positive feedback increases the slope of the rising
edge of the transmission spectrum, as compared to the linear case.
As the frequency extends beyond the resonant mode frequency, the
inner field magnitude in the substrate decreases and the
permittivity goes back towards its linear level, and this negative
feedback keeps the resonant frequency close to the incident field
frequency.
\begin{figure}
\includegraphics[width=80mm]{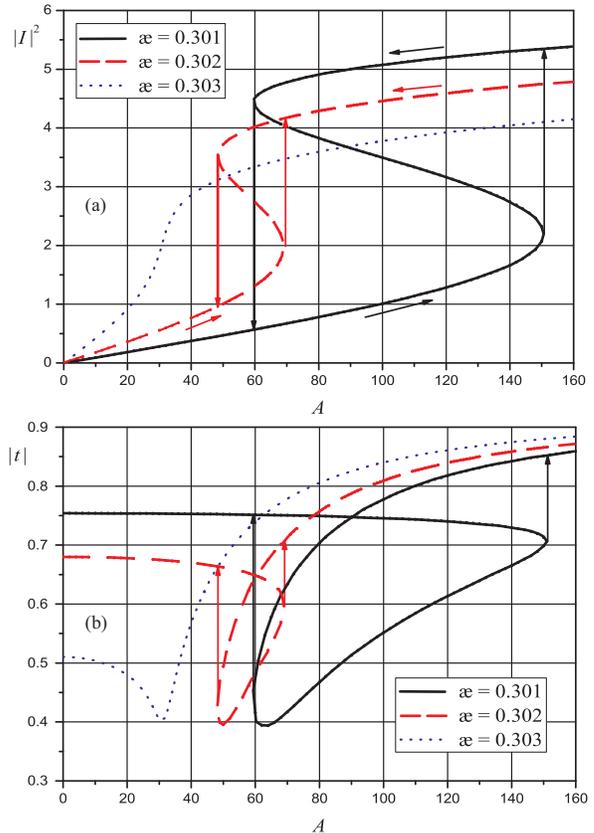}
\caption{(Color online) The square of the current magnitude in a.u.
averaged along split ring (a) and the magnitude of the transmission
coefficient (b) versus the incident field magnitude in the case of
the nonlinear permittivity ($\varepsilon_1=4+0.02i$,
$\varepsilon_2=5\times 10^{-3}$) of the substrate. The arrows
indicate the bistable loop.} \label{fig:nonlinearI&T}
\end{figure}

The curves in Fig.~\ref{fig:nonlinearI&T}a are a plot of the square
of the current magnitude averaged along ASR element versus the
incident field magnitude at fixed frequencies. As an example, let us
consider the incident field frequency $\ae=0.301$ (solid line). As
the incident field amplitude increases, the current magnitude (and
proportionally the inner field intensity) gradually increases along
the bottom branch of the curve, until it reaches about $|I|^2\approx
2$. At this point, the current magnitude jumps to around
$|I|^2\approx 5.5$ due to the instability of the system at the
interior branch of the curve. This transition is shown in the figure
with an arrow directed upward. The incident field magnitude
decreasing results in flowing the square of the current magnitude
along the upper branch of the curve down to $|I|^2\approx 4.5$,
where it drops to a value of about $|I|^2\approx 0.5$. Similarly,
this transition is displayed as an arrow directed downward. The
dramatic current variation from high to low level produces a
switching from high to low transmission (see
Fig.~\ref{fig:nonlinearI&T}b).
\begin{figure}
\includegraphics[width=80mm]{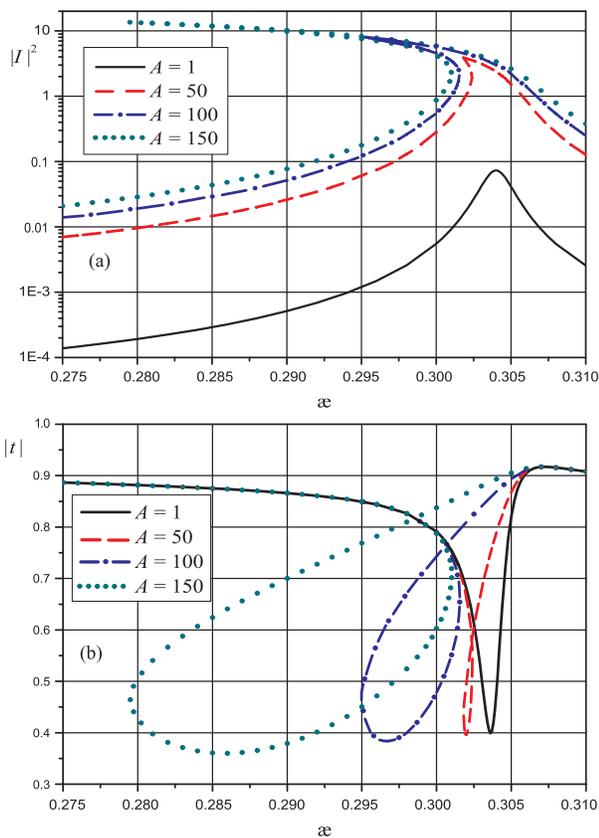}
\caption{(Color online) Frequency dependences of the square of the
current magnitude averaged along split ring (on the logarithmic
scale) in a.u. (a) and the magnitude of the transmission coefficient
(b) in the case of the nonlinear permittivity
($\varepsilon_1=4+0.02i$, $\varepsilon_2=5\times 10^{-3}$) of the
substrate.} \label{fig:fnonlinearI&T}
\end{figure}

The frequency dependence of the transmission coefficient magnitude
manifests also some impressive discontinuous switches to different
values of transmission, as the frequency increases and decreases in
the resonance range for the sufficiently large intensity of the
incident wave. The shifting of the peak of the resonance and the
onset of a bistable transmittivity through the ASR structure is
similar to that of the reflection from a Fabry-Perot cavity
(Fig.~\ref{fig:fnonlinearI&T}a). However, the trapped-mode resonance
is Fano-shaped \cite{sarrazin-2007-bm} rather than the Lorentzian,
as is the characteristic of 1D Fabry-Perot cavities. This Fano
resonance can lead to a peculiar transmission spectra and bistable
behavior \cite{cowan-2003-obi}. In particular, the transmission
resonance of the ASR structure may loop back on themselves
(Fig.~\ref{fig:fnonlinearI&T}b). Such a form of the hysteresis loop
of transmittivity is similar to the behavior of the reflection
spectra of a photonic crystal in the presence of a nonresonant
downstream scattering source in microcavities \cite{cowan-2003-obi};
where the Fano resonances arise due to the interference between two
(or more) different scattering pathways. From the analogy between
these two systems, it can be concluded that the peculiar bistable
loop in the transmission is the result of a superposition of
resonant and nonresonant contributions.

In conclusion, a planar nonlinear metamaterial composed of a metal
array placed on a nonlinear dielectric substrate, which bears a
sharp resonance response by an excitation of a trapped mode due to a
broken symmetry of the pattern is a challenging object for
all-optical switching applications.

This work was completed with the support of the National Academy of
Sciences of Ukraine by the Program "Nanotechnologies and
Nanomaterials", the Project no.~1.1.3.17.

\bibliography{bistability}

\end{document}